\documentclass[bm]{epl}
\usepackage{amssymb}
\title{Effective pair interactions between colloidal particles  at a
  nematic-isotropic interface}
\shorttitle{Colloidal particles at a ni interface}

\author{D. Andrienko\inst{1} \and M. Tasinkevych\inst{2,3} \and S. Dietrich\inst{2,3}}
\shortauthor{D. Andrienko \etal}

 \institute{
  \inst{1} Max-Planck-Institut f\"{u}r Polymerforschung -
             Ackermannweg 10, 55128 Mainz, Germany \\
  \inst{2} Max-Planck-Institut f\"{u}r Metallforschung -
             Heisenbergstr. 3, 70569 Stuttgart, Germany \\
  \inst{3} Institut f\"{u}r Theoretische und Angewandte Physik,
         Universit\"{a}t Stuttgart - Pfaffenwalding 57,
         70569 Stuttgart, Germany
}


\begin{document}

\maketitle

\begin{abstract}
  The Landau-de Gennes free energy is used to study theoretically the
  interaction of parallel cylindrical colloidal particles trapped at a
  nematic-isotropic interface. We find that the effective interaction potential
  is non-monotonic. The corresponding force-distance curves exhibit
  jumps and hysteresis upon approach/separation due to the
  creation/annihilation of topological defects. Minimization results
  suggest a simple empirical pair potential for the effective
  colloid-colloid interaction at the interface.  We propose that the
  interface-mediated interaction can play an important role in
  self-organization and clustering of colloidal particles at such
  interfaces.
\end{abstract}

%######################################################################
\section{Introduction}

Colloidal particles dissolved in a nematic solvent experience
long-ranged
interactions~\cite{ramaswamy.s:1996.a}.
These effective interactions are generated by the distortions of the
liquid crystal director around the particles and result in their
clustering and
self-organization~\cite{loudet.jc:2001.a,poulin.p:1997.a}.
%Due to the presence of topological defects, the interparticle
%interaction is inherently anisotropic: depending on the symmetry of
%the director field around the particles either dipole-dipole or
%quadrupolar-type interactions can be observed~\cite{poulin.p:1997.a}.
%
Upon approaching and at the nematic-isotropic transition, the solvent
forms nematic and isotropic domains, which leads to an additional
interaction of the particles with the interface between the two
phases~\cite{west.jl:2002,andrienko.d:2004a}. The resultant morphology
of the network formed by the particles in colloid-liquid crystal
composites is very sensitive to this
interaction~\cite{west.jl:2002,vollmer.d:2004}. In fact, colloidal
particles can be captured by the nematic-isotropic interface and even
dragged by a {\em moving} interface~\cite{west.jl:2002}. This enables
one to manipulate tiny particles in suspension, which in a more
general context is of technological importance for manufacturing of
$e$-papers and electrophoretic
displays~\cite{zehner.r:1951}, separation of bacterial
species and living
cells~\cite{pohl.ha:1951}, trapping of DNA
and polymer
particles~\cite{washizu.m:1990a}, as
well as growth of photonic
crystals~\cite{yoshinaga.k:1999}.  The structures
of self-assembled colloid layers at {\em free} surfaces of nematic
films can be tuned by their thickness~\cite{smalyukh:2004}.

Recent experiments of dragging colloidal particles by
nematic-isotropic interfaces~\cite{west.jl:2002} demonstrated that a
single colloidal particle is attracted by the interface. This
conclusion is in accordance with theoretical results obtained by
minimizing the corresponding Landau-de Gennes free
energy~\cite{andrienko.d:2004a}. The resulting force on the particle
is non-monotonic and is roughly proportional to the particle radius.
Experimental results~\cite{west.jl:2002} also suggested that the
particles captured by the interface interact with each other
differently than in the pure nematic or isotropic phase, e.g., the
particle-particle interaction depends on the interfacial curvature;
particles which interact only weakly in the bulk can segregate and form
two-dimensional clusters once they are captured by the interface (see
Fig.~\ref{fig:experiment}).

The appearance of additional interface induced features in the
interaction between colloidal particles can be expected from analogous
phenomena at interfaces of simple fluids~\cite{kralchevsky:2000}
which, however, still pose unresolved challenges in
understanding~\cite{wuerger:prl}. In liquid crystals the
situation is even more complicated. The director deformations extend
into the nematic phase, i.~e., in addition to interfacial energies,
the bulk elasticity contributes to the total free energy of the
system.  The effective surface tension (provided it can be introduced
in the first place) depends on the orientation of the director, which
in turn varies along the interface. The boundary conditions for the
director at the particle surfaces give rise to topological defects
accompanying the particles.  In order to minimize the free energy
associated with the director distortions as well as the free energy of
the defect cores, the defects often merge with the isotropic
phase~\cite{andrienko.d:2004a} or with each other.

The combination of the above-mentioned effects enriches the phenomena
significantly but complicates their theoretical understanding
enormously: expressions for the director field, the free energy, or
force-distance profiles can hardly be obtained analytically.
Numerical calculations are also not straightforward because the
interfacial width and the size of the defect cores are normally much
smaller that the size of the colloidal particle, so that there are
several rather different length-scales involved.  Therefore
significant computational efforts involving finite-element methods
with adaptive meshes are required to address even static
problems~\cite{tasinkevych.m:2002.a}.

In the following we study the interaction of two long parallel
cylindrical colloids trapped at a (initially) flat nematic-isotropic
interface.  We find that their effective interaction is rather
complex: the force-distance profiles can exhibit jumps and hysteresis
upon approach or separation. The particles can either repel or
attract each other depending on the material parameters and the
geometry. Finally, we propose a simple effective pair potential which
can be exploited to study clustering and self-organization of colloids
at NI interfaces.

%######################################################################
\section{Landau--de Gennes free energy}
\label{sec:theory}

\begin{figure}
\twofigures[height=3cm]{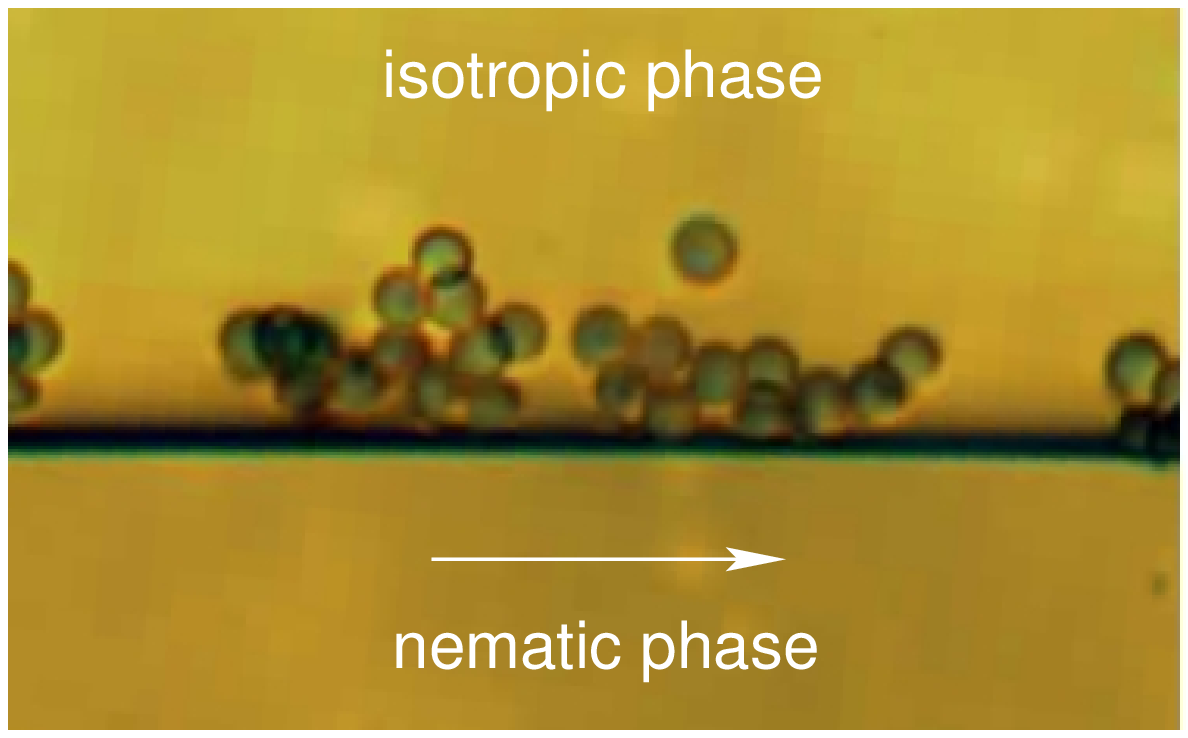}{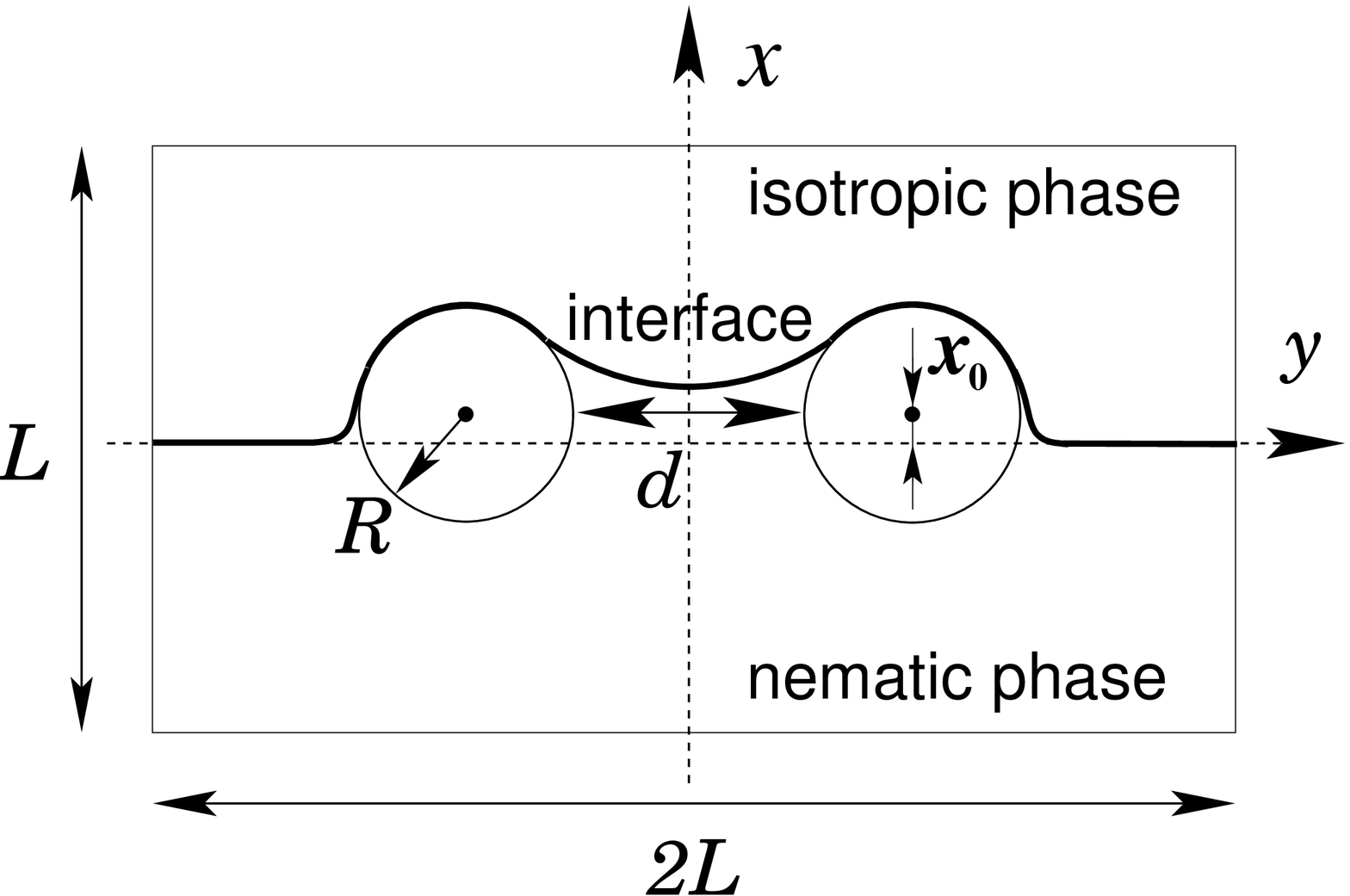}

  \caption{Aggregates of $16\, \mu \rm m$ polymer particles captured by a
  nematic-isotropic interface. The director is parallel to the nematic-isotropic
  interface and lies  within the plane of the figure. In the experiment the interface 
  moves in the upward direction and drags the colloids.
  Courtesy of A.~Glushchenko, J.~L.~West, G.~Liao, and Yu.~Reznikov,
  Liquid Crystal Institute, Kent, Ohio, USA.}
  \label{fig:experiment}

  \caption{Cross-section of two parallel cylindrical particles trapped at a
    nematic-isotropic interface at $x=0$. The system is translationally
    invariant in $z$-direction.}
  \label{fig:geometry}

\end{figure}

%The geometry we study is shown in Fig.~\ref{fig:geometry}. 
Two identical colloidal particles, each of which we take to be a long
cylinder of radius $R$ with the symmetry axis parallel to the $z$
axis, are immersed into a nematic liquid crystal at a separation $d$,
see Fig.~\ref{fig:geometry}. The nematic order parameter tensor at
the boundaries is fixed in a such a way that without the colloids a
flat nematic-isotropic interface is formed at $x=0$. The order
parameter at the top wall is fixed to zero; the absolute value of the
order parameter at the bottom wall is fixed to the bulk order
parameter of the nematic phase at two-phase coexistence. The director
orientation at the bottom wall is also fixed and corresponds to the
preferred director anchoring at the nematic-isotropic interface, i.e.,
parallel or perpendicular to the interface. 
The axes of the particles are positioned at $x=x_0$. In the course of the free
energy minimization $x_0$ adjusts itself such that that the $x-$component
of the force exerted on the particles vanishes.
%We use the phenomenological description of the orientational order
%based on the symmetric traceless second-rank tensor order parameter
%$Q_{ij}$. 

The system is characterized by the Landau-de Gennes free
energy~\cite{degennes.pg:1995.a}
\begin{eqnarray}
F = \int dV \Bigl[ a(T-T^*)Q_{ij}Q_{ji} - bQ_{ij}Q_{jk}Q_{ki} +
c(Q_{ij}Q_{ji})^2 +\frac{L_1}{2} Q_{ij,k}Q_{ij,k}
+\frac{L_2}{2}Q_{ij,j}Q_{ik,k} \Bigr ] \label{eq:nem_free_en},
\end{eqnarray}
where summation over repeated indices is implied and the comma
indicates the spatial derivative.  The positive constants $a,b,c$ are
assumed to be temperature independent, and $T^*$ is the supercooling
temperature of the isotropic phase.  The constants $L_1$ and $L_2$ are
related to the Frank-Oseen elastic constants $K_{11} = K_{33} =
9Q_b^2(L_1 + L_2/2)/2$ and $K_{22} = 9Q_b^2L_1/2$ and $Q_b$ is the
bulk nematic order parameter. The sign of $L_2$ determines the
preferred orientation of the director at the nematic-isotropic
interface. $L_2>0$ ($L_2<0$) favors planar (perpendicular)
anchoring.  We introduce the dimensionless
temperature $\tau = 24ca(T-T^*)/b^2$. The bulk nematic phase is stable
for $\tau < \tau_{\rm NI} = 1$ with a degree of orientational order
given by $Q_b = b(1+\sqrt{1-8\tau/9})/8c$; $Q_b(\tau>1)=0$.

We consider both rigid homeotropic and finite anchoring boundary
conditions at the colloidal surfaces.  In the latter case the surface
free energy density is taken as~\cite{nobili.m:1992} $f_s
=\frac{1}{2}w {\rm Tr} { \left( {\bm Q} - {\bm S} \right)^2}$ where
$w$ is the anchoring coefficient and $S_{\alpha \beta} = S \left(
  e_{\alpha} e_{\beta} - \delta_{\alpha \beta}/3 \right)$ is the
preferred tensor order parameter at the colloid surface with ${\bm e}$
as the unit vector along the easy axis direction.

%%%%%%%%%%%%%%%%%%%%%%%%%%%%%%%%%%%%%%%%%%%%%%%%%%%%%%%%%%%%%%%%%%%%%%%%%%
%\subsection{Parameters}
For the constants entering the free energy density
(Eq.~(\ref{eq:nem_free_en})) we use typical values for a nematic
compound 5CB ~\cite{coles.hj:1978}: $a = 0.044 \times
10^6 \,{\rm J/m^3K}$, $b = 0.816 \times 10^6 \,{\rm J/m^3}$, $c = 0.45
\times 10^6 \,{\rm J/m^3}$, $L_1 = 6 \times 10^{-12} \,{\rm J/m}$,
$T^* = 307 \,{\rm K}$. The choices for $L_2$ will be given below.
 The nematic-isotropic transition temperature
for 5CB is $T_{NI} =308.5 \,{\rm K}$. At coexistence the nematic
coherence length (i.e., the thickness of the nematic-isotropic
interface) is $\xi = \left( 24 L_1 c/b^2 \right)^{1/2} \approx 10\,
\rm nm $ which sets the smallest length-scale of our description.

The equilibrium distribution of the tensor order parameter $Q_{ij}$ is
obtained by minimizing the free energy
functional~(\ref{eq:nem_free_en}) numerically using finite elements
with adaptive meshing. The area $L \times 2L$ is triangulated and the
functions $Q_{ij}$ are linearly interpolated within each triangle.

\section{Planar interfacial anchoring}

\begin{figure}
\twofigures[height=5.7cm]{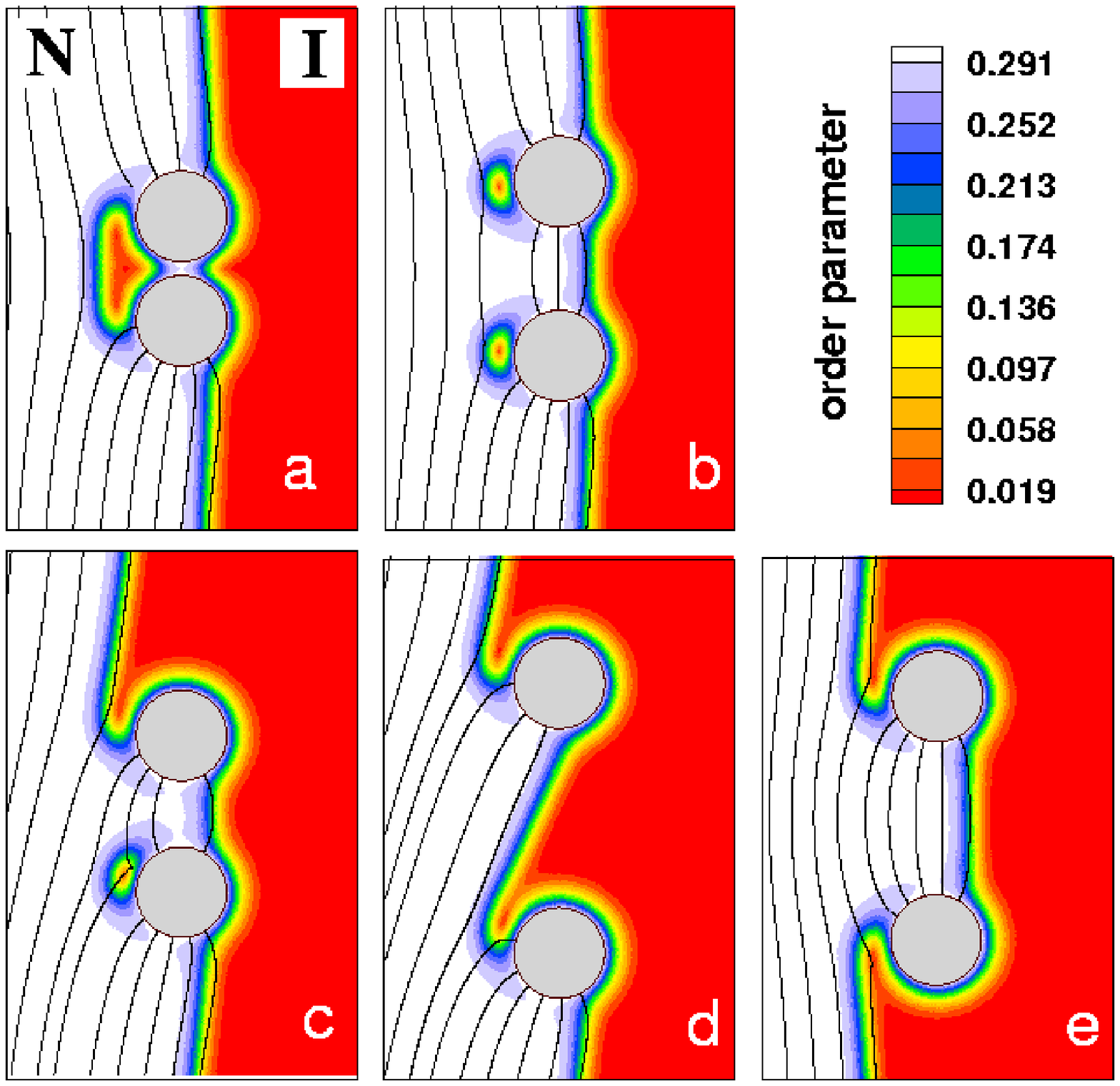}{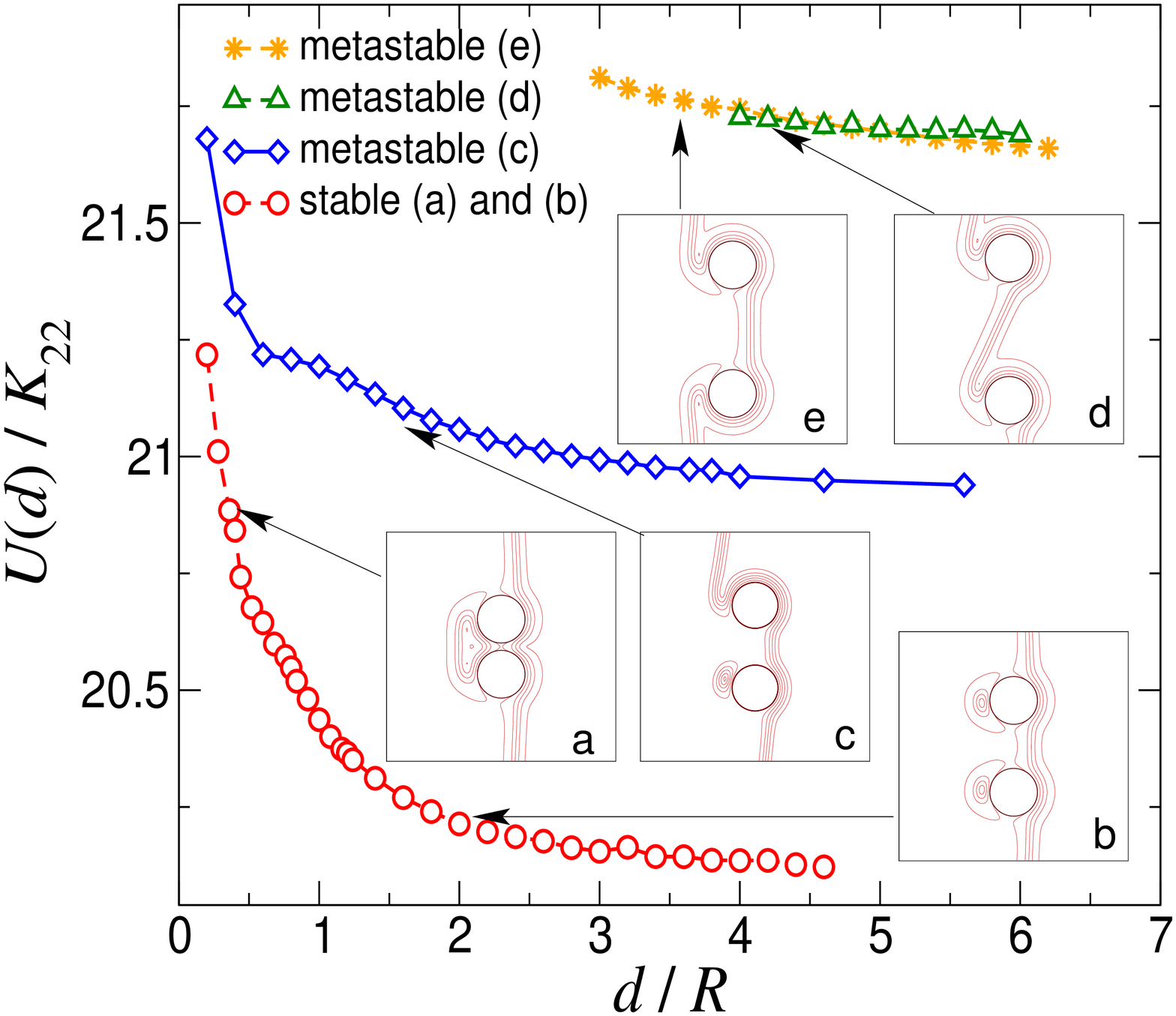}
\caption{
  Order parameter and director maps for stable \{a,b\} and metastable
  \{c,d,e\} configurations. Particle radius $R/\xi=5$; elastic constants
  favor planar anchoring at the nematic-isotropic interface. Solid lines
  indicate the local director orientation.
  There is translational symmetry along the $z$ axis and the director lies
  in the $xy$ plane.}
\label{fig:config_planar}

\caption{
  Effective pair potential $U(d)$ per unit length for planar director anchoring at the NI
  interface and homeotropic anchoring at the colloid surface,
   for both stable and metastable configurations
  with $R/\xi=5$. The insets depict the corresponding
  configurations by iso-order-parameter lines.}
\label{fig:energy_planar}

\end{figure}

For this case we chose the anisotropy of the elastic constants as
$L_2/L_1 = 2$, favoring director alignment parallel to the NI
interface and in the plane $z=0$. We use rigid homeotropic boundary conditions at the
colloidal surfaces. Upon minimizing the free energy we have
identified two stable orientational configurations. At
sufficiently large separations $d$ both colloids are
accompanied by a defect of strength $1/2$
(Fig.~\ref{fig:config_planar}(b)).
If the particles move closer, the two
defects merge and form a single, but rather extended, defect
positioned between the colloids (Fig.~\ref{fig:config_planar}(a)).
We have also observed three metastable configurations. Here either one
of the defects, Fig.~\ref{fig:config_planar}(c), or even both defects,
Fig.~\ref{fig:config_planar}(d) and (e), merge with the isotropic
phase: the system is trying to reduce its free energy by annihilating
the defects. However, the interface has to bend and almost wrap the
particles in order to reach the defects. This of course increases the
interfacial energy, which is roughly proportional to the area of the
interface.
For small separations, among the metastable configurations only the
one shown in Fig.~\ref{fig:config_planar}(c) is possible: due to
strong anchoring at the particle surfaces the nematic phase forms a
bridge between the particles, preventing the isotropic phase from
reaching also the second defect. For larger separations, the isotropic
phase can extend partially around the colloids and merge with the
second defect. This can occur asymmetrically, as shown in
Fig. \ref{fig:config_planar}(d), or symmetrically, see
Fig. \ref{fig:config_planar}(e).

The effective pair potential $U(d)$ defined as the free energy per
length of the colloidal particles is presented in
Fig.~\ref{fig:energy_planar}. In all cases the particles repel each
other. 
%This repulsion is rather strong at small distances, due to the
%strong homeotropic anchoring at the particle surfaces, and decays rapidly
%for large $d$. 
The branch of $U(d)$ which corresponds to the {\em
  stable} configurations, reveals practically no  breaks in slope,
even though at $d/R\approx 0.4$ the director field around the particles changes abruptly
between the configurations (a) and (b). However, if one of the metastable
configurations is present in the system, transitions between different
branches can occur giving rise to jumps and hysteresis in the
force-distance curves upon particle approach/separation.

%%%%%%%%%%%%%%%%%%%%%%%%%%%%%%%%%%%%%%%%%%%%%%%%%%%%%%%%%%%%%%%%%%%%%%%%%%%
\section{Homeotropic anchoring at the interface}

\begin{figure}
\twofigures[height=5.7cm]{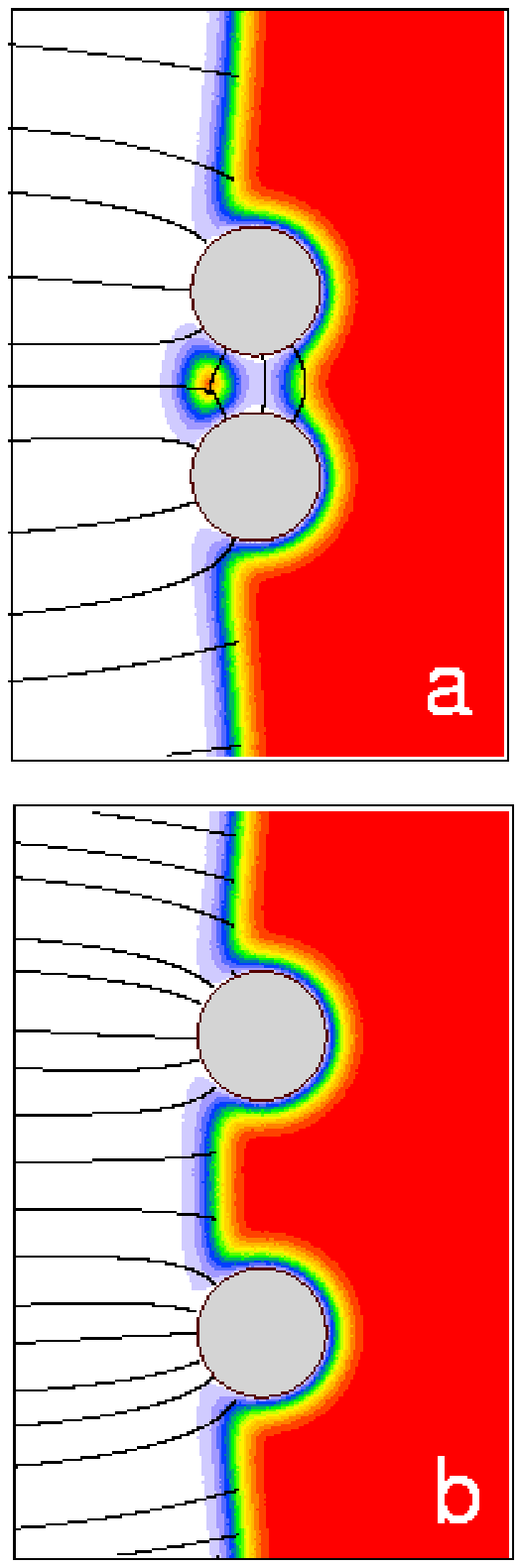}{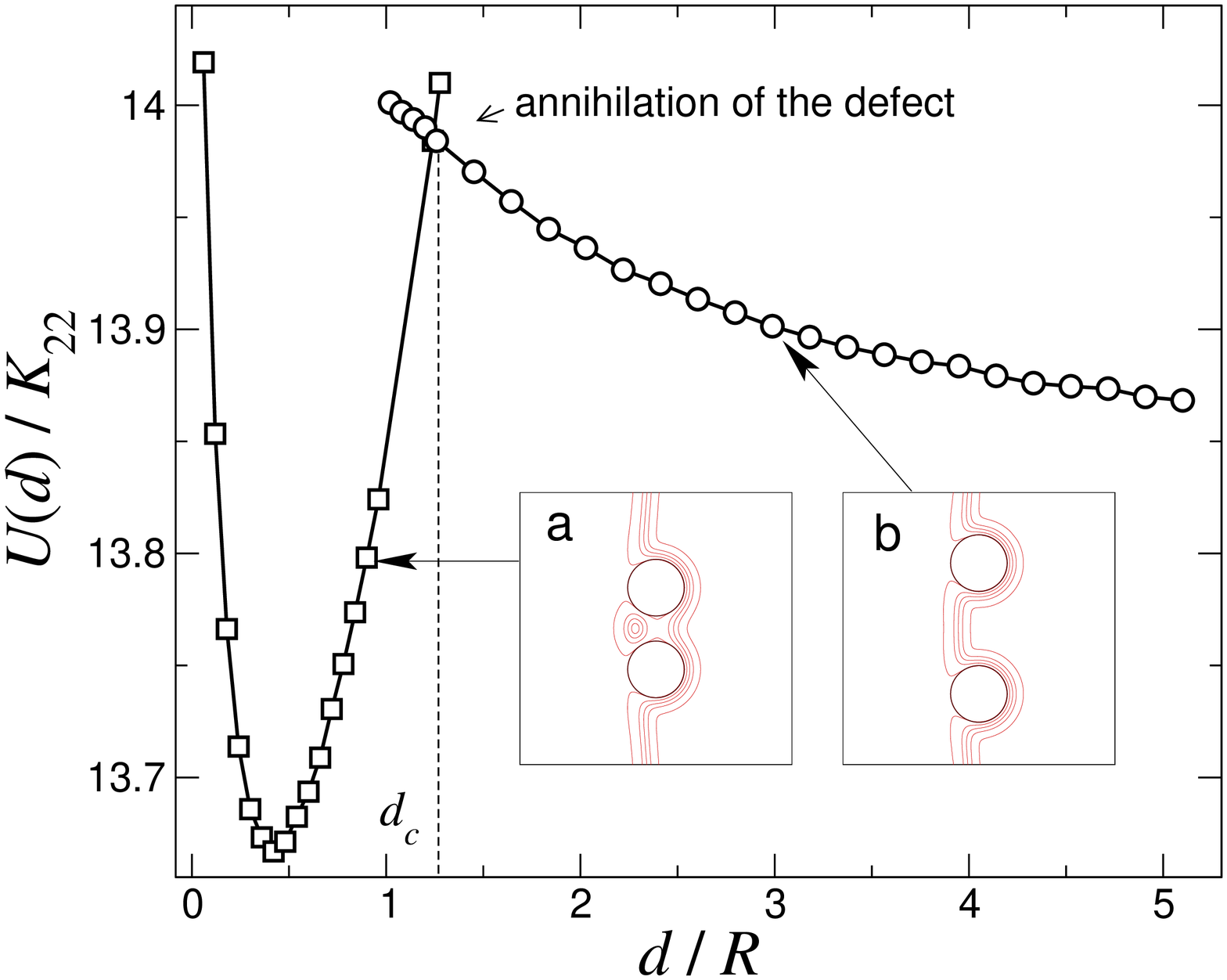}

\caption{
  Director and order parameter maps for homeotropic anchoring at the
  NI interface corresponding to elastic constants $L_2 = - 1/2 L_1$
  and particle radius $R/\xi=5$. Configuration (a) is stable at short
  distances, (b)  at large distances.  }
\label{fig:config_hom}

\caption{
  Effective pair potential $U(d)$ per unit length for homeotropic anchoring at the NI
  interface, $L_2 = -1/2 L_1$, homeotropic anchoring at the colloid surface,
   and $R/\xi=5$.  The configuration
  with the defect results in strong attraction at short distances.
  There is repulsion at large distances. The insets characterize the
  corresponding configurations by iso-order-parameter lines (see
  Fig.~\ref{fig:config_hom}). For $d<d_c$ configuration (b) is metastable.}
\label{fig:energy_hom}

\end{figure}

The choice $L_2/L_1 = -1/2$ for the anisotropy of the elastic
constants favors director alignment perpendicular to the NI interface in the plane $z=0$.
In this case we have found only two, stable configurations. The first
one, containing a defect, is shown in Fig.~\ref{fig:config_hom}(a). It
occurs at small separations, where the nematic bridge between the
particles forces the director to be parallel to the interface.  The
other configuration has no defects, because the nematic bridge
with director orientation parallel to the interface
disappears and the defect merges with the isotropic phase.

The corresponding effective potential $U(d)$ is shown in
Fig.~\ref{fig:energy_hom}. As in the previous case, the particles
repel each other at small separations where there are large director
distortions in the inter-particle region. However, contrary to the
case with  planar anchoring, there is a strong attraction between the
particles for $d/R \lesssim 1$, where the configuration with the defect
is stable. The branch without the defect results in weak repulsion, as
for planar anchoring.

%%%%%%%%%%%%%%%%%%%%%%%%%%%%%%%%%%%%%%%%%%%%%%%%%%%%%%%%%%%%%%%%%%%%%%
\section{Empirical expression for the effective potential}

Minimization of the free energy is a computationally expensive task,
which cannot be used, for example, to calculate the forces on the
particles each time their positions are advanced using molecular
dynamics or Monte-Carlo algorithms for their kinetics. Use of effective potentials is a
typical solution to this problem. 
%
%It can be efficiently used at
%practically any length-scale: in the coarse-grained models [..],
%atomistic simulations, or {\em ab initio} calculations [..].  
%
%Once the potential is known, one can consider a two-dimensional motion
%of particles trapped by the interface interacting via an additional
%effective force derived from this potential.

In order to find such an appropriate potential we have calculated the free
energy per unit length for particles of different radii, $R/\xi
\approx 2.5-12.5$, corresponding to $R \approx 0.025-0.125\, \mu \rm
m$. Since colloids are rather small, the finite director anchoring at
the particle surfaces has to be taken into account. We have used a
value appropriate for 5CB on a polymer
surface~\cite{andrienko.d:2002.a}, $w\xi/L_1 = 0.822$; for this energy
the dimensionless anchoring parameter $\eta=wQ_b^2R/K_{22}$ is of the
order of $1$.

%This corresponds to the values of the dimensionless anchoring
%parameter $\eta=wQ_b^2R/K_{22}$ ranging from $\approx 0.463$ to
%$\approx 2.315$, for $R/\xi$ from $2.534$ to $12.671$. 

\begin{figure}
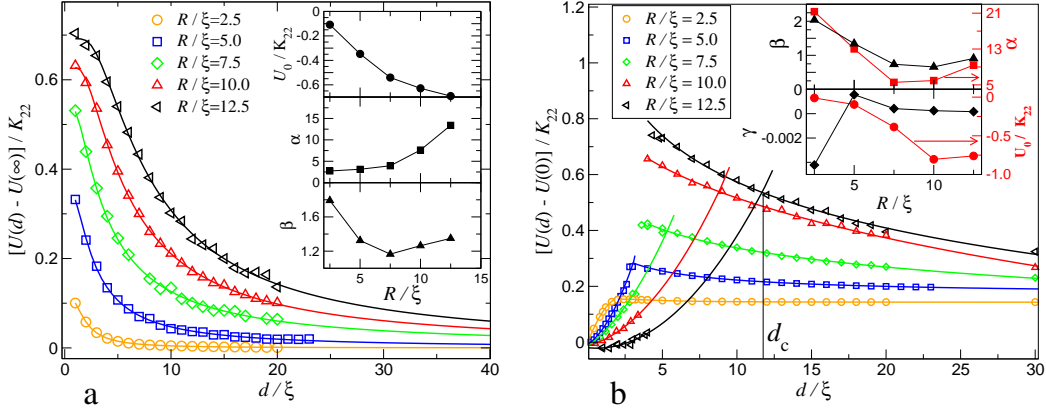

  \twoimages[height=5.4cm]{figures/energy_paral_fit.eps}
  {figures/energy_perp_fit.eps}
\caption{
  Effective pair potential as a function of the distance $d$ between
  colloids for different colloid radii $R$: (a) planar and (b) homeotropic
  anchoring of the director at the interface and
  finite anchoring at the colloid surface. Symbols: minimization
  results; solid lines: fit to the empirical equation~(\ref{eq:fit}).
  Inset illustrates the dependence of the fitting parameters on $R$. 
  $\xi = 10\, \rm nm$,  $K_{22} = 2.6 \times 10^{-12}\, \rm J/m$.
   $U_a/K_{22} \approx 6.8-26.5$ corresponding to $R/\xi \approx 2.5 - 12.5$.  }
\label{fig:energy_fit}
\end{figure}

Minimization results for both planar and homeotropic anchoring of the
director at the interface are shown in Fig.~\ref{fig:energy_fit} (a)
and (b). 
%As before, planar anchoring results in a
%purely repulsive interparticle force whereas there is strong attraction
%at short distances in case of homeotropic anchoring. 
The strong repulsion close to contact ($d=0$) has become smoother for
planar and disappeared for homeotropic anchoring at the interface, due
to the finite director anchoring at the colloid surfaces.
It turns out that in both cases the following shape of the potential
fits the numerical data very well:
\begin{equation}
U(d) = \left\{
\begin{array}{cc}
U_a \left[ 1+\gamma(d/\xi)^2 \right] , & 0 \le d \le d_c \\
U(\infty) + U_0 \Bigl (\exp 
\left[ -\alpha \left( {d}/{\xi}  \right)^{-\beta} \right] - 1\Bigr), & d > d_c,
\end{array}
\right.
\label{eq:fit}
\end{equation}
where $\alpha, \beta, \gamma$ are (positive) fitting parameters. The
first part of the potential ($0 \le d \le d_c$) describes a simple
harmonic-type attraction between the particles, which we observed for
the {\em homeotropic} anchoring; for {\em planar} interfacial
anchoring $d_c \equiv 0$. In the case of homeotropic anchoring $d_c$ is a
critical distance at which the defect annihilates (see
Figs.~\ref{fig:config_hom} and \ref{fig:energy_hom}).

%Note that all
%fitting parameters ($U_a, U_0, \alpha, \beta, d_c$) depend on the
%particle radius $R$.

We have no analytical support for the functional form of the effective
potential (\ref{eq:fit}). However, qualitatively, the effective
interaction of the colloidal particles in a bulk nematic phase is known to
exhibit a power-law decay at large
distances~\cite{ramaswamy.s:1996.a}.
In our case, the leading term of the expansion for $d/\xi \rightarrow \infty$ 
is indeed $\sim(d/\xi)^{-\beta}$. On the
other hand, if two colloids are sufficientlly close, so that
the regions where the {\em order parameter} changes from their surface
to the bulk value start to overlap, and where the defects (if present)
interact with each other and the
particles~\cite{tasinkevych.m:2002.a}, the functional form of
the potential changes. Normally, the distortions of the order
parameter relax exponentially (similar to the variations of the order
parameter through the nematic-isotropic interface) resulting in an
exponentially varying strength of the interaction. This is also
captured qualitatively by the potential given in Eq.~(\ref{eq:fit}).

The fitting curves are also shown in Fig.~\ref{fig:energy_fit} (a) and
(b). For planar anchoring $\beta \approx 1.5$; for homeotropic
anchoring $\beta \approx 1.0$, i.~e., in both cases the interaction
potential is long-ranged. In case of homeotropic anchoring there is a
critical size of colloidal particles (of the order of interfacial
width) below which the discontinuous structural transition between the
configurations with and without the defect disappears. Similar
phenomena, named capillary bridging, can be observed in granular
materials, liquid crystals, and binary
mixtures~\cite{stark.h:2004.a}. In all cases, one of the phases
(otherwise metastable in the bulk) condenses out and forms a bridge
between the particles. This transition is discontinuous and ends in a
critical point for small particle radii. In our case, the nematic
phase located between the isotropic phase and the defect forms a
bridge between the particles and results in their attraction.

\section{Conclusions}
In conclusion, within the Landau-de Gennes formalism we have studied
the effective pair interaction between two cylindrical parallel
colloids trapped at the nematic-isotropic interface. This pairwise
interparticle force depends sensitively on the radii of colloids, the
orientation of the director at the NI interface, and the anchoring
strength, i.e., the nematic elasticity is an essential ingredient of
the interaction. The formation or annihilation of defects
influences the effective pair potential.  It also changes the
repulsive character of the interaction at large separations to an
attractive one at intermediate or small separations.  Therefore, one
should expect that many-body interactions are relevant to describe
collective colloidal ordering at the interface. Another implication of
our work is that the interface is able to rearrange the nucleation
centers of, for example, phase-separating polymer network and help to
design composite materials of extraordinary properties.

\acknowledgments 

We thank to M.~Deserno, A.~Glushchenko, K.~Kremer, G.~Liao,  M. Oettel,
Yu.~Reznikov, and J.~L.~West for useful discussions.

%\bibliography{letter}

\end{document}